\documentclass[aps,prl,reprint,superscriptaddress]{revtex4-2}
\usepackage{graphicx}
\usepackage{amsmath,amssymb}
\usepackage{times} 
\usepackage{float}

\usepackage{url}
\begin{document}
\title{Universality and anisotropy of the Photonic Urbach Tail}

\author{M. Men\'endez}
\affiliation{Instituto de Ciencia de Materiales de Madrid (ICMM), Consejo Superior de Investigaciones Cient\'ificas (CSIC), Sor Juana In\'es de la Cruz 3, 28049 Madrid, Spain}

\author{Lan Hoang Mai}
\affiliation{Quantum Science and Engineering Program, University of Delaware, Newark, Delaware 19716, USA}

\author{Nazifa Tasnim Arony}
\affiliation{Department of Materials Science and Engineering, University of Delaware, Newark, Delaware 19716, USA}

\author{Henry Carfagno}
\affiliation{Department of Materials Science and Engineering, University of Delaware, Newark, Delaware 19716, USA}

\author{Lauren N McCabe}
\affiliation{Department of Materials Science and Engineering, University of Delaware, Newark, Delaware 19716, USA}

\author{Joshua M. O. Zide}
\affiliation{Department of Materials Science and Engineering, University of Delaware, Newark, Delaware 19716, USA}

\author{Cefe L\'opez}
\affiliation{Instituto de Ciencia de Materiales de Madrid (ICMM), Consejo Superior de Investigaciones Cient\'ificas (CSIC), Sor Juana In\'es de la Cruz 3, 28049 Madrid, Spain}

\author{Matthew F. Doty}
\affiliation{Quantum Science and Engineering Program, University of Delaware, Newark, Delaware 19716, USA}
\affiliation{Department of Materials Science and Engineering, University of Delaware, Newark, Delaware 19716, USA}

\author{P. D. Garc\'ia}
\email{pd.garcia@csic.es}
\affiliation{Instituto de Ciencia de Materiales de Madrid (ICMM), Consejo Superior de Investigaciones Cient\'ificas (CSIC), Sor Juana In\'es de la Cruz 3, 28049 Madrid, Spain}

\date{\today}

\begin{abstract}
Disorder in photonic crystals and waveguides creates states inside the photonic band gap. These states are often described as Lifshitz tails despite exhibiting energy distributions inconsistent with Lifshitz statistics near the band edge. Here we show that in photonic-crystal waveguides with intentionally engineered anisotropic disorder, the band-edge tail accessible experimentally follows an Urbach law universally, with cumulative statistics $F(\Delta)=\exp[-(\Delta/\alpha)^\beta]$, where $\Delta$ is the spectral detuning from the band edge, and an exponent $\beta \approx 1$ independent of disorder strength and orientation. In contrast to Lifshitz behavior, the density of states is maximal at the band edge and decays into the gap. Crucially, we find that the Urbach energy $\alpha$ is anisotropic, with a pronounced directional splitting and qualitatively different scaling for disorder parallel and perpendicular to the waveguide axis. These conclusions are supported by quantitative agreement between optical measurements of GaAs photonic-crystal waveguides and full-vector simulations. The anisotropic Urbach energy emerges as a sensitive probe of disorder--mode coupling and a practical metric to characterize structural disorder in photonic devices.
\end{abstract}

\maketitle

Disorder is an essential element of transport in general and of wave propagation in any realistic material system, from electronic conduction in solid-state matter to optical propagation in photonic devices. In a photonic crystal waveguide, disorder-induced states -- typically discussed in the context of Anderson localization -- extend into the photonic bandgap~\cite{Topolancik2007,Garcia2017,Hughes2005,Patterson2009,Kuramochi2005,Mazoyer2009}. The spectral distribution of these states broadens band edges by generating a nonzero density of states (DOS) within the bandgap, which is typically labeled a Lifshitz tail~\cite{Huisman2012,Garcia2013,Garcia2010,Savona2011,VascoHughes2017}. However, the Lifshitz regime describes a DOS that vanishes at the band edge and increases as one moves deeper into the gap by an energy distance $\Delta > 0$, according to $\rho(\Delta) \propto \exp\left[-A_d/\Delta^{d/2}\right]$, where $d$ is the spatial dimensionality of the system and $A_d > 0$~\cite{Lifshitz1963, Lifshitz1964}. In contrast, the tails observed experimentally show the opposite behavior: the DOS \emph{decreases} into the gap following $\rho(\Delta) \propto \exp(-\Delta/E_U)$-the hallmark of the Urbach law identified originally in semiconductor absorption~\cite{Urbach1953,John1986}. Indeed, John \emph{et al.}~\cite{John1986} showed that when disorder has finite spatial correlation the entire observable near-edge tail follows Urbach statistics. Moreover, recent work on compositionally-disordered photonic crystals has described the observed band-edge broadening qualitatively as Urbach-like~\cite{Lee2018}. A central open question is whether these Urbach statistics retain a universal form under anisotropic structural disorder, and whether such anisotropy is encoded in the Urbach energy itself.

\begin{figure}[t!]
  \includegraphics[width=\columnwidth]{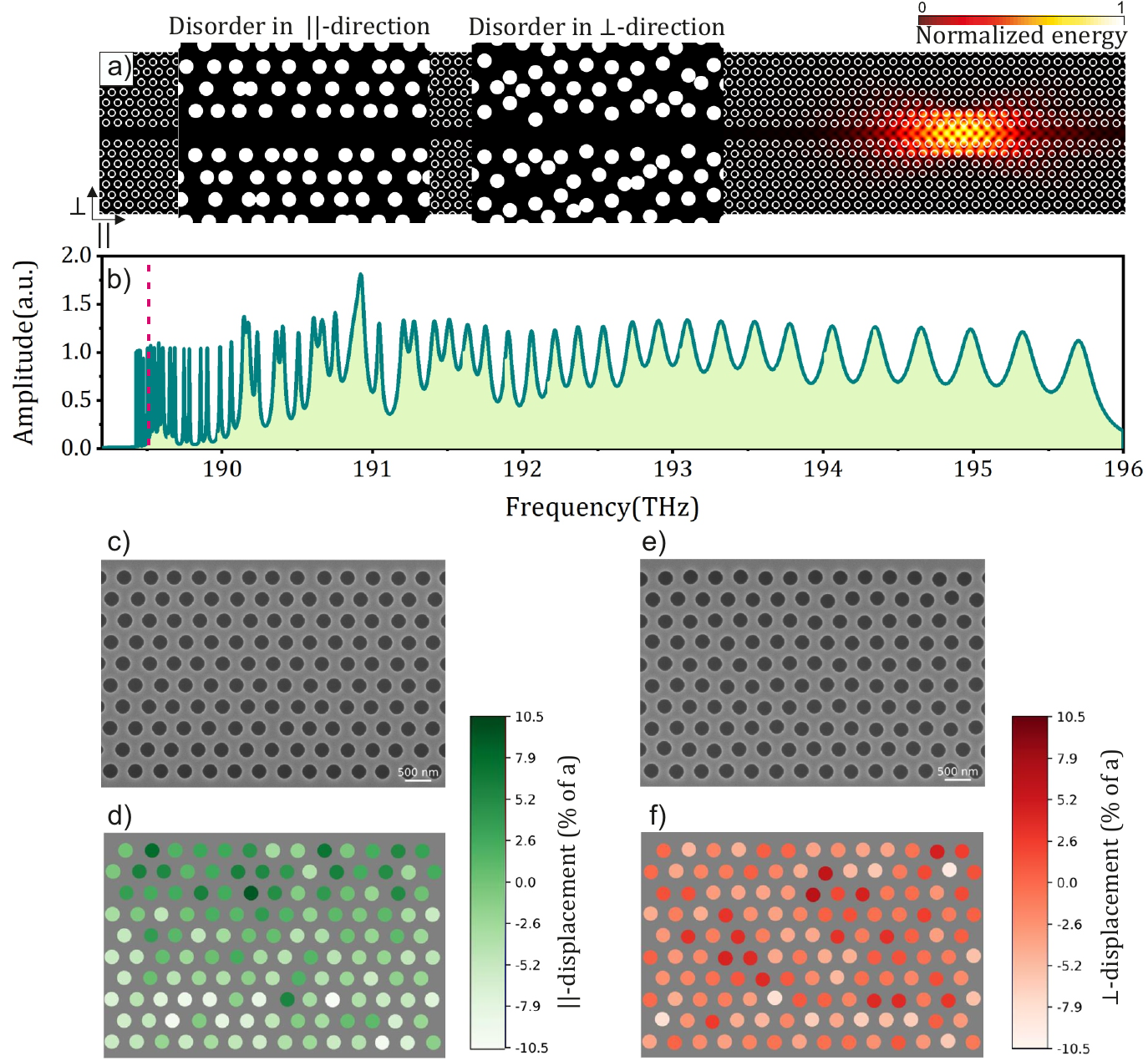}
\caption{\textbf{Photonic-crystal waveguides with controlled anisotropic disorder.}
(a) Waveguide geometry (triangular lattice of air holes with $a = 500$~nm the lattice constant) with hole displacements along $\parallel$ and $\perp$ directions, and the intensity profile of an Anderson-localized mode.
(b) Simulated spectrum showing Fabry--Perot-like modes and cutoff frequency at $189.5~\text{THz}$ (dashed line) when disorder is introduced in the $\parallel$ direction ($\sigma_\parallel = 0.01a$).
(c,d) SEM image and positional disorder map for $\parallel$ disorder ($\sigma_\parallel = 0.03a$).
(e,f) SEM image and positional disorder map for $\perp$ disorder ($\sigma_\perp = 0.03a$).
Color scale indicates hole displacement from ideal lattice position. Strong contrast in (d) and (f) confirms disorder is predominantly along the intended direction.}
  \label{fig1}
\end{figure}

\begin{figure*}[t]
  \includegraphics[width=\textwidth]{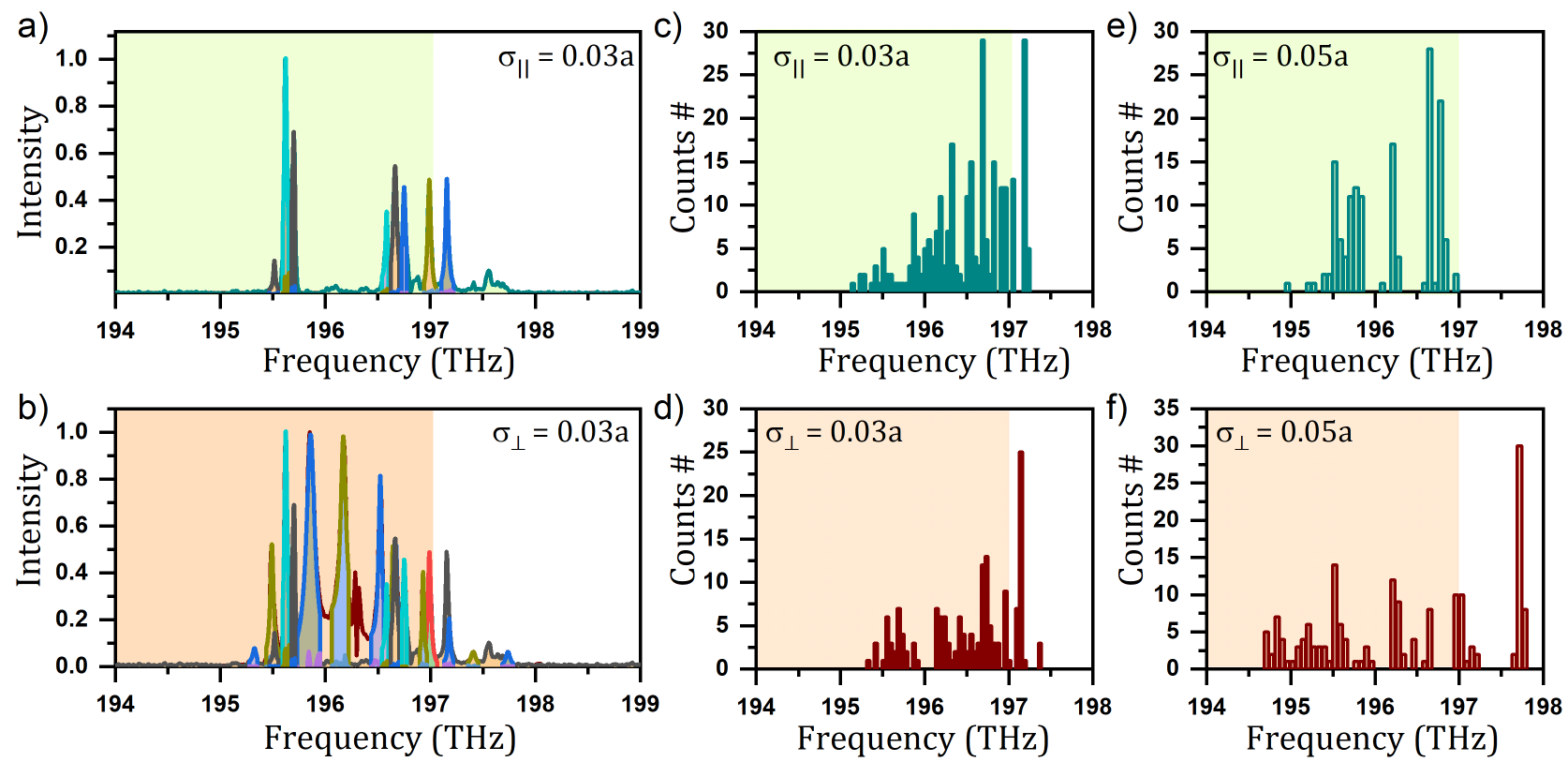}
 \caption{\textbf{Experimental observation of Urbach tails.} (a),(b) Representative optical reflection spectra from GaAs photonic-crystal waveguides with intentional disorder along $\parallel$ (a) and $\perp$ (b) at $\sigma \approx 0.03a$. Each curve shows an individual Anderson-localized resonance detected via evanescent fiber coupling. (c)--(f) Eigenfrequency histograms for all detected localized modes; shaded regions indicate the band gap. At weak disorder $\sigma = 0.03a$ (c,d), $\parallel$  and $\perp$ disorder distributions are similar due to comparable intrinsic isotropic disorder ($\sigma \simeq 0.007a$). At stronger disorder $\sigma = 0.05a$ (e,f), clear anisotropy emerges: the $\perp$ disorder tail extends $\sim 1$~THz deeper into the gap, confirming $\alpha_\perp > \alpha_\parallel$.}
  \label{fig2}
\end{figure*}

Here we show that photonic band-edge tails in disordered photonic-crystal waveguides follow Urbach statistics universally, even under strongly anisotropic structural disorder. We observe this result through optical evanescent coupling experiments using a tapered fiber to probe the near-field response of GaAs waveguides fabricated with controlled directional disorder parallel ($\parallel$) and perpendicular ($\perp$) to the waveguide axis, and confirm these results numerically using two-dimensional finite-element simulations. We fit the cumulative frequency distribution of localized states to the stretched-exponential form $F(\Delta) = \exp[-(\Delta/\alpha)^\beta]$, where the scale parameter $\alpha$ corresponds to the Urbach energy $E_U$ when $\beta = 1$. We extract a robust tail exponent $\beta \approx 1$, characteristic of the Urbach regime, across all disorder amplitudes and orientations. While the functional form of the tail remains universal, the Urbach energy shows a clear anisotropy: the exponent $\beta$ is invariant, but the energy scale $\alpha$ exhibits strong directional dependence, with $\alpha_{\parallel} \neq \alpha_{\perp}$. Moreover, $\alpha_{\parallel}$ remains nearly constant with disorder strength, whereas $\alpha_{\perp}$ increases markedly. This reveals that the Urbach energy provides a sensitive directional probe of disorder--mode coupling.

Figure~\ref{fig1} introduces the photonic-crystal waveguide system and our approach to controlled anisotropic disorder. Panel (a) illustrates the waveguide geometry---a triangular lattice of air holes with a missing row defining the guiding channel---along with a representative Anderson-localized mode that forms near the band edge. Panel (b) shows the simulated transmission spectrum of a structure when disorder is introduced in the $\parallel$ direction ($\sigma_\parallel = 0.01a$, with a lattice constant $a = 500$~nm), exhibiting Fabry--Perot-like oscillations above the cutoff frequency $\omega_c = 189.5$~THz; below this frequency, propagation is forbidden and disorder-induced localized states appear. To realize anisotropic disorder experimentally, we fabricate GaAs photonic-crystal waveguides using electron-beam lithography, generating mask files in which each hole position is displaced from its ideal lattice site by a random offset drawn from a Gaussian distribution with standard deviation $\sigma_{\parallel}$ or $\sigma_{\perp}$ applied exclusively along the $\parallel$ or $\perp$ direction (see Supplementary Information for fabrication details). All samples also contain intrinsic fabrication disorder---isotropic random positional fluctuations with $\sigma \simeq 0.007a$ (see Supplementary Information for this evaluation). Panels (c--f) in Fig.~\ref{fig1} present scanning electron microscope images together with real-space maps of the measured positional disorder, extracted by fitting hole positions and referencing them to an ideal lattice. The displacement maps confirm that the fabricated disorder is strongly directional: panel (d) shows large displacements along $\parallel$ with minimal $\perp$ component, while panel (f) shows the opposite.

In order to probe the optical DOS, we perform optical reflection measurements on samples covering intentional disorder amplitudes from $\sigma = 0.01a$ to $0.05a$. Figure~\ref{fig2} presents the experimental data. Panels (a) and (b) show representative reflection spectra for samples with intentional disorder along the direction $\parallel$ to the waveguide at $\sigma_{\parallel} \approx 0.03a$ (panel a) and along the $\perp$ direction at $\sigma_{\perp} \approx 0.03a$ (panel b). Each colored curve represents an individual Anderson-localized resonance detected at a different spatial position along the waveguide. We fit each resonance with a Lorentzian lineshape to extract the resonant frequency $\omega_i$ and linewidth $\gamma_i$.

Panels (c--f) in Fig.~\ref{fig2} present eigenfrequency histograms compiled from all detected localized modes. At weak intentional disorder of $\sigma = 0.03a$ [panels (c) and (d)], the frequency distributions for $\parallel$ disorder and $\perp$ disorder are nearly identical, showing no signs of anisotropy. We attribute this to the intrinsic isotropic disorder ($\sigma \simeq 0.007a$) present in all samples being comparable to this weak intentional anisotropic disorder, which masks directional signatures. However, at stronger intentional disorder of $\sigma = 0.05a$ [panels (e) and (f)], clear anisotropy emerges: the DOS brought about by $\perp$ disorder extends significantly deeper into the gap than that originating from $\parallel$ disorder, with the tail reaching $\sim 1$~THz lower in frequency, confirming $\alpha_\perp > \alpha_\parallel$ qualitatively. This threshold behavior shows that Urbach tails probe the total disorder: directional anisotropy becomes observable experimentally only when intentional disorder exceeds intrinsic levels by a factor of $\sim 2$--$3$.

\begin{figure}[t!]
  \includegraphics[width=\columnwidth]{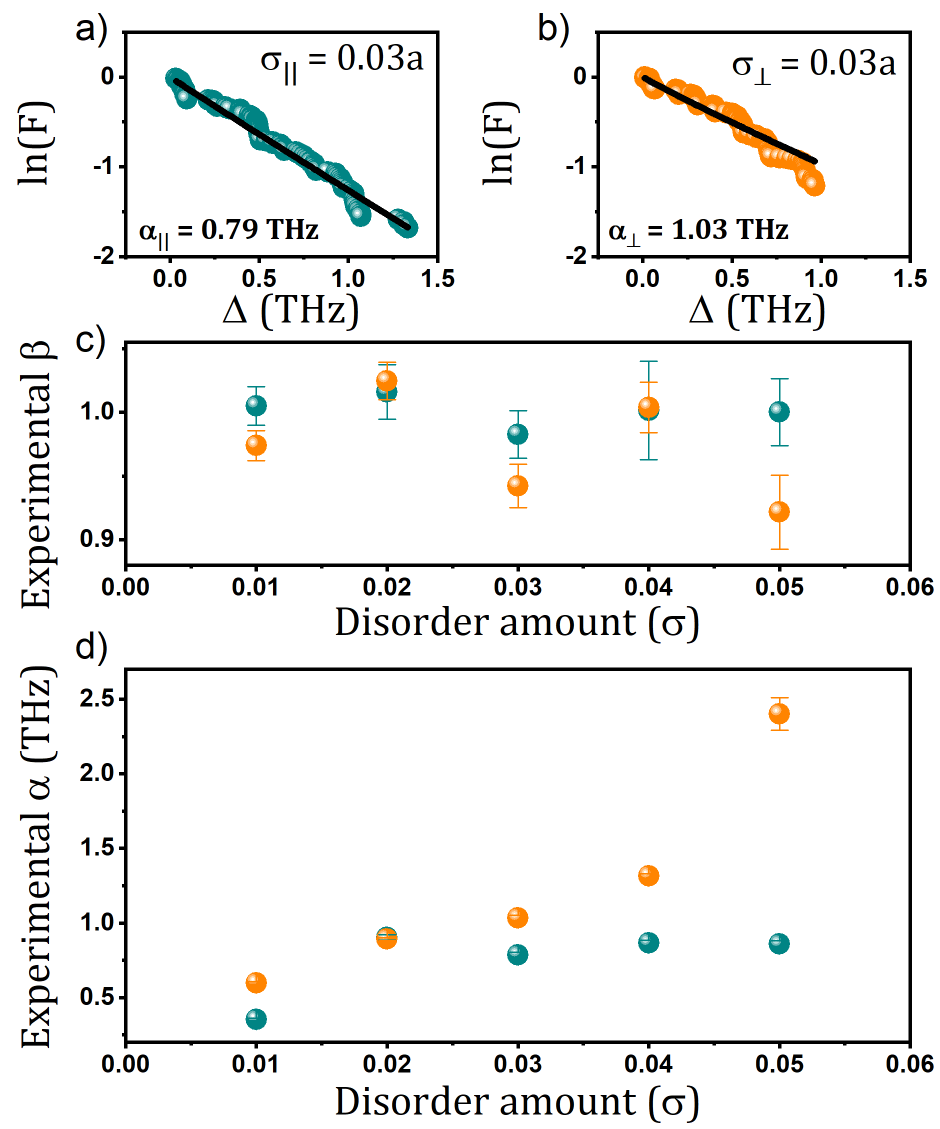}
  \caption{\textbf{Anisotropy of the Urbach tail.} (a,b) Urbach fits to experimental data: $\ln F$ versus $\Delta$ for $\parallel$ disorder (a, teal) and $\perp$ disorder (b, orange) at $\sigma = 0.03a$. Black lines show fits to $F(\Delta) = \exp[-(\Delta/\alpha)^\beta]$. Steeper decay for $\perp$ disorder confirms $\alpha_\parallel < \alpha_\perp$. (c) Experimental Urbach exponent $\beta$ versus disorder amplitude for both directions, showing $\beta \approx 1$ across all cases, confirming universal Urbach behavior. (d) Experimental Urbach energy $\alpha$ versus disorder amplitude. $\alpha_\parallel$ (teal) remains nearly flat while $\alpha_\perp$ (orange) increases linearly with $\sigma_{\perp}$, in quantitative agreement with simulations (Fig.~\ref{fig4}f).}
  \label{fig3}
\end{figure}

Figure~\ref{fig3} presents the quantitative Urbach analysis of the experimental data. We fit the stretched-exponential form $F(\Delta) = \exp[-(\Delta/\alpha)^\beta]$ to the experimental frequency distributions, extracting the Urbach exponent $\beta$ and scale parameter $\alpha$ for each disorder configuration. Panels (a) and (b) show the Urbach fits for $\sigma = 0.03a$: we plot $\ln F$ versus $\Delta$ for $\parallel$ disorder (panel a, teal) and $\perp$ disorder (panel b, orange). The excellent linearity confirms that the experimental spectral tails follow the Urbach form. The fitted slopes yield $\alpha_\parallel < \alpha_\perp$, as confirmed by simulations below. Panel (c) shows the extracted Urbach exponent $\beta$ versus disorder amplitude for both directions. Remarkably, despite the presence of intrinsic disorder and the complexities of real fabricated structures, $\beta \approx 1$ for all measured configurations. This confirms that Urbach tail statistics ($\rho(\Delta) \propto \exp(-\Delta/\alpha)$) apply universally, independent of the direction or magnitude of the disorder. We note that the experimental frequency extraction is performed by fitting Lorentzian lineshapes to reflection spectra at discrete fiber positions; as a result, modes extracted from the same spectrum share a common baseline and wavelength calibration, which can introduce small correlations (visible as mild clustering) without affecting the global tail statistics. Panel (d) presents the experimental Urbach energies versus disorder amplitude. The experimental data reveal a key result: $\alpha_\parallel$ (teal) remains nearly flat across disorder strengths while $\alpha_\perp$ (orange) increases approximately linearly with $\sigma_{\perp}$. As shown below, finite-element simulations (Fig.~\ref{fig4}f) reproduce these experimental findings quantitatively, validating the robustness of Urbach tails as a probe of anisotropic disorder coupling in photonic structures. The absence of a clear directional splitting at the lowest disorder levels is consistent with the presence of an intrinsic isotropic fabrication background, with anisotropic Urbach scaling emerging only once the introduced intentional disorder exceeds this baseline.

\begin{figure*}[t!]
  \includegraphics[width=\textwidth]{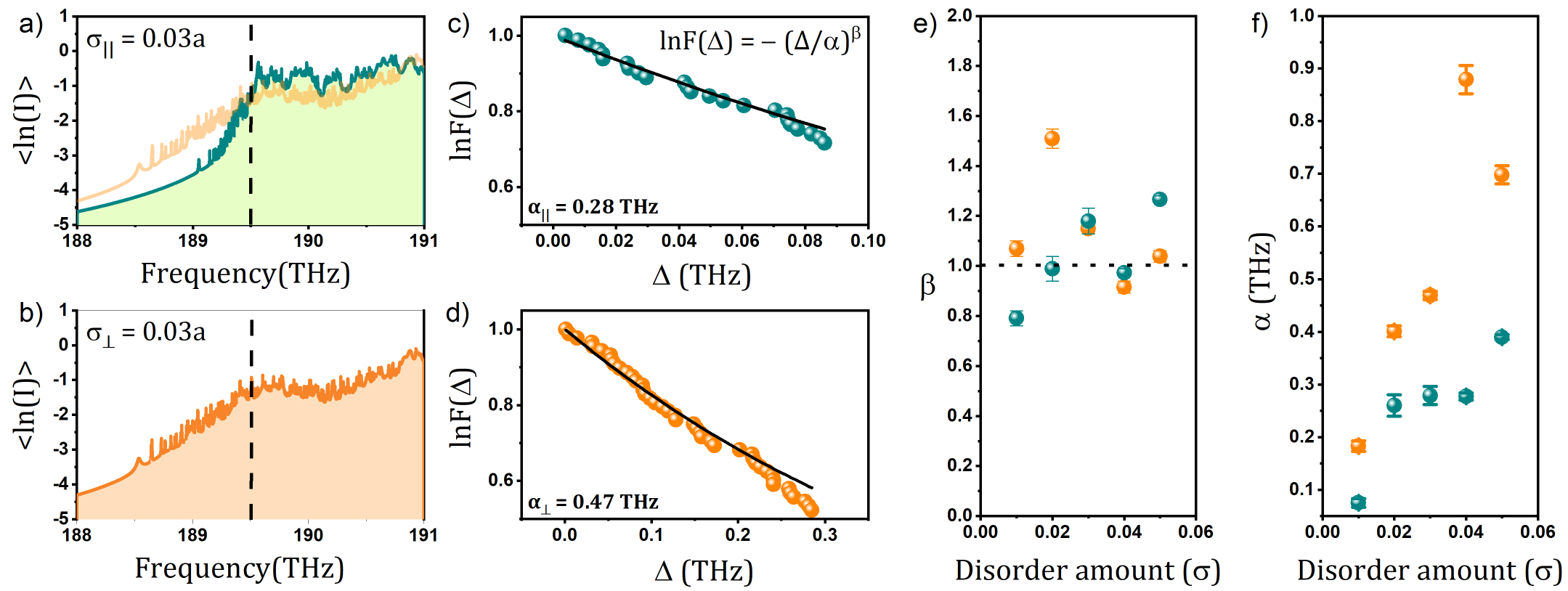}
\caption{\textbf{Numerical analysis of the Urbach tails.}
  (a),(b) Ensemble-averaged spectra for $\parallel$ (a, teal) and $\perp$ (b, orange) disorder at $\sigma = 0.03a$, showing sub-edge states below the band edge at $\omega_c = 189.5~\text{THz}$. The faint orange overlay curve in (a) shows the perpendicular disorder direction for comparison; the $\perp$ disorder tail extends deeper into the gap. (c),(d) Urbach analysis: $\ln F(\Delta)$ versus detuning $\Delta$ for $\parallel$ disorder (c, teal) and $\perp$ disorder (d, orange), with $F$ the cumulative DOS. The almost linear decay confirms exponential band-edge tails; the steeper slope for $\perp$ disorder implies $\alpha_\parallel < \alpha_\perp$.
  (e) Urbach exponent $\beta$ versus disorder amplitude, with $\beta \approx 1$ for all $\sigma$ and both directions (dashed line indicates $\beta = 1$), consistent with Urbach-type near-edge behavior. (f) Urbach energy $\alpha$ versus disorder amplitude. $\alpha_\parallel$ (teal) is nearly constant, while $\alpha_\perp$ (orange) grows with $\sigma_{\perp}$, revealing directional sensitivity.}
  \label{fig4}
\end{figure*}

Figure~\ref{fig4} presents the statistical analysis of the disorder-induced spectral tails from numerical simulations. For each disorder configuration, we compute eigenfrequencies using finite-element simulations and construct the cumulative distribution $F(\Delta) = N^{-1}\sum_i \Theta(\Delta_i - \Delta)$, where $N$ is the total number of modes, $\Delta_i = \omega_c - \omega_i$ is the detuning of mode $i$ from the band edge, and $\Theta$ is the Heaviside step function. We fit the stretched-exponential form $F(\Delta) = \exp[-(\Delta/\alpha)^\beta]$ with both $\alpha$ and $\beta$ as free parameters. Panels (a) and (b) compare ensemble-averaged spectra for $\parallel$ and $\perp$ disorder at $\sigma = 0.03a$: the $\perp$ disorder tail extends significantly deeper into the gap. Panels (c) and (d) show the Urbach analysis: plotting $\ln F$ versus $\Delta$ reveals excellent linearity, confirming exponential tail behavior.

The key results appear in panels (e) and (f) in Fig.~\ref{fig4}. Panel (e) shows the fitted exponent $\beta$ versus disorder amplitude: remarkably, $\beta \approx 1$ for all disorder strengths and both directions, confirming universal Urbach-type behavior rather than the Lifshitz form ($\beta \approx -1/2$) expected for rare deep-gap states. This is consistent with the fact that true Lifshitz states lie deep in the gap and are effectively inaccessible in finite structures, whereas Urbach tails arise from collective band-edge broadening and dominate the observable statistics. Panel (f) reveals striking anisotropy: while $\alpha_\parallel$ (teal) remains nearly constant with increasing $\sigma_{\parallel}$, $\alpha_\perp$ (orange) grows approximately linearly with $\sigma_{\perp}$. This directional dependence reflects how structural fluctuations along different crystal axes couple to the guided Bloch mode: transverse ($\perp$) disorder modulates the waveguide width and couples more strongly to the band-edge mode than longitudinal ($\parallel$) disorder along the propagation direction.

In summary, we show that disorder-induced spectral tails at photonic band edges follow an exponential Urbach form within the experimentally accessible near-edge regime, with cumulative statistics $F(\Delta) = \exp[-(\Delta/\alpha)^\beta]$ and a robust exponent $\beta \approx 1$ that is insensitive to disorder strength and isotropic. Crucially, the $\alpha$ scale parameter---the Urbach energy $E_U$---is not universal but encodes directional information: $\alpha_{\parallel} \neq \alpha_{\perp}$, and the two components exhibit qualitatively different disorder scalings, directly reflecting how structural fluctuations couple to Bloch modes along the waveguide axis versus perpendicular to it. Thus the Urbach energy emerges as a quantitative, experimentally accessible probe of anisotropic disorder--mode coupling at band edges. More broadly, these findings establish Urbach physics as a natural framework for exponential tail statistics in structured wave systems. The directional Urbach tensor provides a practical, experimentally accessible metric to quantify anisotropic scattering and the onset of localization in any wave system where structural anisotropy couples to propagating modes---from photonic crystals to cold-atom lattices and acoustic metamaterials---and poses new questions about the role of disorder in crystals and hyperuniform structures.

\begin{acknowledgments}
This work was supported by the Spanish Ministry of Science, Innovation and Universities via the national project PID2024-158832NB-C21 (PSYNC), by the HORIZON-EIC-2022 Pathfinder project NEUROPIC (Grant No. 101098961), and by the Spanish MICIU Severo Ochoa program for Centres of Excellence through Grant CEX2024-001445-S. This research was partially supported by the US NSF through the University of Delaware Materials Research Science and Engineering Center (DMR-2011824) and the EPSCoR RII program (2217786).
\end{acknowledgments}

\clearpage
\onecolumngrid

\section*{Supplementary Information}

\setcounter{figure}{0}
\setcounter{table}{0}
\renewcommand{\thefigure}{S\arabic{figure}}
\renewcommand{\thetable}{S\Roman{table}}

\section{THEORETICAL BACKGROUND: URBACH VS LIFSHITZ TAILS}

This section clarifies the distinction between Urbach and Lifshitz tails in disordered photonic systems. A critical confusion in the photonics literature has been the mislabeling of near-band-edge disorder-induced states as ``Lifshitz tails'' when they actually exhibit Urbach tail behavior. We analyze the stretched-exponential form $F(\Delta) = \exp[-(\Delta/\alpha)^\beta]$ using the cumulative distribution function $F(\Delta)$ where $\beta$ is treated as a free fitting parameter and $\alpha = E_U$ is the Urbach energy with units of frequency. The sign and magnitude of the exponent $\beta$ distinguishes between the two regimes: $\beta \approx +1$ for Urbach tails near the band edge, while for Lifshitz tails deep in the gap the exponent is negative with $\beta \approx -1/2$, corresponding to $\rho(\Delta) \propto \exp(-A_d/\sqrt{\Delta})$. Critically, these two regimes exhibit opposite monotonic behavior of the density of states as a function of detuning from the band edge.

\begin{figure}[b!]
  \centering
  \includegraphics[width=0.95\textwidth]{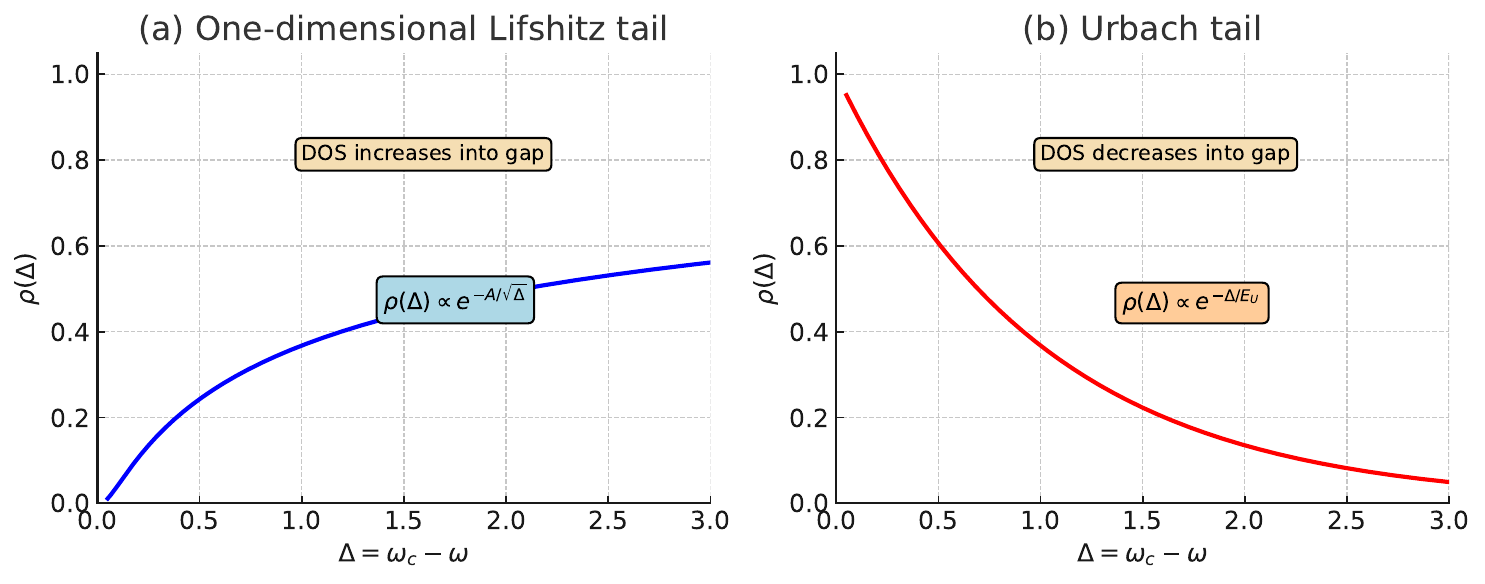}
  \caption{Comparison between the qualitative behavior of
(a) a Lifshitz tail and (b) an Urbach tail. Panel (a) illustrates a \emph{one-dimensional} Lifshitz form $\rho(\Delta)\propto \exp[-A\,\Delta^{-1/2}]$, where the exponent varies as an inverse power of the detuning and the density of states therefore \emph{increases} as one moves deeper into the gap. This captures the key qualitative feature of the Lifshitz regime predicted for uncorrelated disorder, whose general $d$-dimensional form is
$\rho(\Delta)\sim\exp[-A\,\Delta^{-d/2}]$. Panel (b) shows an Urbach tail
$\rho(\Delta)\propto\exp(-\Delta/E_U)$, where the exponent is linear in
$\Delta$ and the DOS \emph{decreases} into the gap. The purpose of the figure is to highlight the opposite monotonicity of Lifshitz and Urbach behavior. Only the Urbach regime is observed in the band-edge spectra of photonic-crystal waveguides, consistent with the correlated-disorder theory of John \emph{et al.}~\cite{john1986}.}
  \label{fig:urbach_vs_lifshitz}
\end{figure}

\subsection{The Urbach tail: disorder-induced band-edge broadening}

The Urbach tail was first observed by Urbach in 1953~\cite{urbach1953} in the optical absorption edge of disordered solids. Urbach found that the absorption coefficient near the band edge follows an exponential law with a characteristic energy scale $E_U$ called the Urbach energy, which quantifies disorder-induced broadening. For the density of states below the band edge with detuning $\Delta = \omega_c - \omega > 0$, this corresponds to an exponential decay $\rho(\Delta) \propto \exp(-\Delta/E_U)$. The Urbach tail describes band-edge smearing due to disorder, where the density of states decreases exponentially as one moves away from the band edge into the gap. The physical origin is the broadening of the band edge by local potential fluctuations caused by structural disorder. In Urbach's original work on thermal disorder, the Urbach energy was found to be proportional to temperature, $E_U \approx kT$, establishing temperature as the relevant energy scale. The connection between Urbach tails and disorder-induced localized states in electronic systems was established by John et al.~\cite{john1986} in their theory of electron band tails. More recently, Lee et al.~\cite{lee2018} confirmed Urbach tail behavior in photonic band-tail states of compositionally disordered photonic crystals, explicitly noting the similarity to semiconductor absorption edges.

\subsection{The Lifshitz tail: rare-region states deep in the gap}

In the original theory of Lifshitz for uncorrelated disorder in $d$
dimensions~\cite{Lifshitz1964SI}, the density of states deep inside the
gap obeys the asymptotic form
\[
\rho(\Delta)\sim \exp[-A\,\Delta^{-d/2}],
\]
where $\Delta=\omega_c-\omega>0$ is the detuning from the band edge and
$A$ is a constant determined by the disorder amplitude and the spatial
dimensionality. The key feature is the \emph{exponent}:
the argument of the exponential varies as an \emph{inverse power}
$\Delta^{-d/2}$. Because this exponent decreases as $\Delta$ increases,
the DOS actually \emph{rises} when moving deeper into the gap—opposite
to the Urbach tail, where the exponent is proportional to $\Delta$ and
the DOS therefore \emph{decays} into the gap. Moreover, as shown by
John \emph{et al.}~\cite{john1986}, when the disorder has a finite
spatial correlation length the Lifshitz asymptotic is pushed into an
experimentally inaccessible deep-gap region, and the entire observable
near-edge tail follows an Urbach tail.

\subsection{Distinguishing the two regimes}

The key distinction between Urbach and Lifshitz tails lies in the behavior of the density of states as one moves into the gap. In the Urbach regime, characteristic of near-band-edge states, the density of states decreases exponentially into the gap following $\rho(\Delta) \propto \exp(-\Delta/E_U)$. In contrast, the Lifshitz regime describes deep-gap states where the density of states follows $\rho(\Delta) \propto \exp(-A_d/\sqrt{\Delta})$, exhibiting the opposite behavior where the density increases deeper into the gap. Most importantly, these two regimes exhibit opposite monotonic behavior: the Urbach tail shows monotonically decreasing density of states as one moves into the gap, while the Lifshitz tail shows monotonically increasing density of states deeper into the gap. This opposite spectral dependence provides the clearest physical diagnostic to distinguish between these two fundamentally different disorder regimes, as illustrated in Fig.~\ref{fig:urbach_vs_lifshitz}.

A critical clarification is needed regarding the photonics literature. Several prominent papers have incorrectly labeled band-edge tails as ``Lifshitz tails'' despite observing decreasing density of states moving into the gap, which is characteristic of Urbach behavior. For example, Huisman et al.~\cite{huisman2012} measured the DOS near the band edge of a photonic crystal waveguide and observed a tail that ``slowly decays away from the band edge,'' yet referred to it as ``the Lifshitz tail known from solid-state systems.'' Similarly, Garc\'ia et al.~\cite{garcia2013} used the term ``Lifshitz tail'' to describe the spectral width of Anderson-localized modes near the waveguide cutoff, despite the characteristic exponential decrease of the DOS into the gap.

\section{NUMERICAL SIMULATIONS}

\subsection{Computational methodology}

\begin{figure*}[t!]
  \centering
  \includegraphics[width=\textwidth]{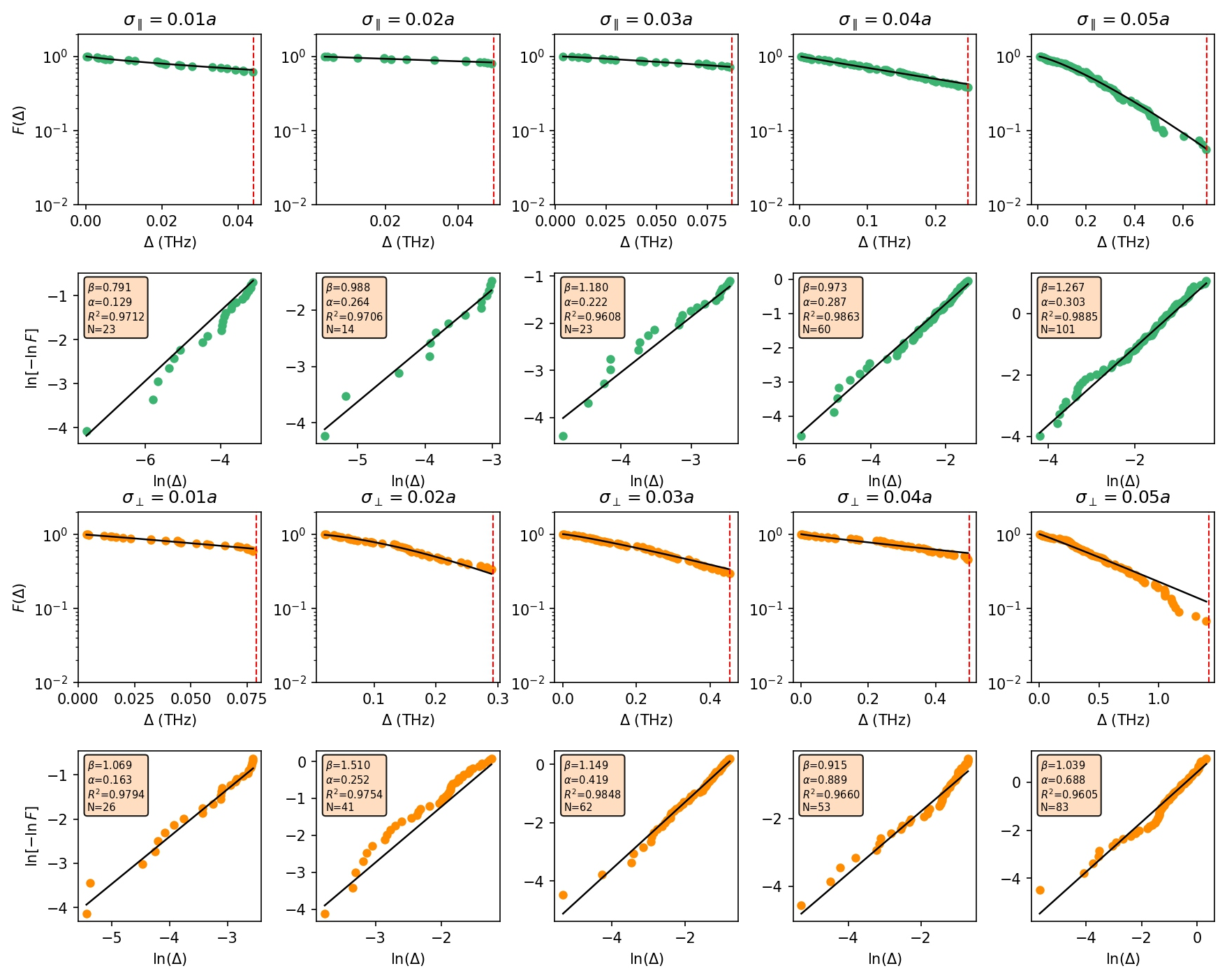}
  \caption{Numerical fitting analysis for directional disorder. The layout consists of 4 rows and 5 columns, where columns correspond to disorder amplitudes ($\sigma_\parallel$ or $\sigma_\perp = 0.01a$ through $0.05a$). Rows 1-2 show $\parallel$ disorder results: row 1 displays $F(\Delta)$ versus $\Delta$ (THz), and row 2 displays the linearized fit $\ln[-\ln F]$ versus $\ln\Delta$. Rows 3-4 show $\perp$ disorder results with the same layout. Colored points (teal for $\parallel$ disorder, orange for $\perp$ disorder) show the optimized fitting region satisfying $0.05 < F(\Delta) < 0.95$ and $\Delta < \Delta_{\mathrm{max}}$. Black lines show the linear fits, which yield slope = $\beta$ and intercept = $-\beta\ln\alpha$, from which we obtain the Urbach energy $\alpha = E_U$ via $\alpha = \exp(-\mathrm{intercept}/\beta)$. Vertical dashed red lines mark $\Delta_{\mathrm{max}}$ for each case. The fitted parameters ($\beta$, $\alpha$, $R^2$, and $N$) are shown in the insets of the linearized plots. The excellent linearity confirms that the stretched-exponential form accurately describes the numerical spectral tails, and the fitted exponent values $\beta \approx 1$ confirm Urbach tail behavior.}
  \label{fig:numerical_fits}
\end{figure*}

We employ the finite element method using COMSOL Multiphysics to solve the two-dimensional eigenvalue problem for the electromagnetic wave equation in the frequency domain. The photonic-crystal waveguide geometry consists of a triangular lattice of air holes with lattice constant $a = 500$~nm and hole radius $r = 0.29a$. The structure is modeled in two dimensions using an effective refractive index $n_{\mathrm{eff}} = 2.265$ to reproduce the TE-like guided band of the real GaAs slab. The computational domain extends $N_x = 180$ unit cells along the propagation direction ($x$), resulting in a system length of 90~$\mu$m. In the transverse direction, the waveguide cross-section contains $2N_y = 8$ rows of holes on each side of the missing waveguide row (see Fig.~1(a) in the main text). This is a standard photonic-crystal waveguide cross-section and the Anderson-localized modes that form the band-edge tail are strongly confined within the central few rows, with negligible field amplitude at the outer rows. As a result, edge effects from the finite transverse extent of the simulated structure do not influence the calculated eigenfrequencies.

Disorder is implemented as random position fluctuations of all holes in the lattice. Critically, to study anisotropic disorder effects, each hole is displaced exclusively either along the $\parallel$ direction or along the $\perp$ direction, creating directionally biased disorder. For $\parallel$ disorder, each hole position is displaced by a random amount drawn from a Gaussian distribution with zero mean and standard deviation $\sigma_\parallel$, applied only along the $\parallel$ coordinate. Similarly, for $\perp$ disorder, displacements with standard deviation $\sigma_\perp$ are applied only along the $\perp$ coordinate. This creates an anisotropic disorder tensor with principal components $\sigma_\parallel^2$ or $\sigma_\perp^2$, allowing us to probe how structural fluctuations along different crystallographic directions couple to the photonic band structure. We scan disorder amplitudes $\sigma = 0.01a$, $0.02a$, $0.03a$, $0.04a$, and $0.05a$ for both $\parallel$ and $\perp$ directions.

Perfectly matched layers are applied at both waveguide terminations to model an open system and suppress spurious reflections. The eigenvalue solver computes complex eigenfrequencies $\lambda$ near a target frequency of 189 THz, close to the waveguide cutoff. From each eigenvalue $\lambda$, we extract the real frequency $\omega = \mathrm{Im}(-\lambda)/(2\pi)$ and the Q-factor $Q = \mathrm{Im}(-\lambda)/(2\mathrm{Re}(\lambda))$. Spurious modes with unphysically low Q-factors below 200 are discarded. For each physical eigenmode, the detuning from the band edge is computed as $\Delta_i = \omega_c - \omega_i$ where $\omega_c = 189.5$ THz is the waveguide cutoff frequency. For each disorder amplitude and direction, we perform simulations over an ensemble of approximately 10 disorder realizations. After collecting eigenmodes from all disorder realizations, the cumulative distribution is constructed as $F(\Delta) = N^{-1} \sum_{i=1}^N \Theta(\Delta_i - \Delta)$, where $N$ is the total number of physical modes across all realizations and $\Theta(x)$ is the Heaviside step function.

\subsection{Fitting procedure with free exponent}

For empirical analysis of the spectral tail, we construct the cumulative distribution function $F(\Delta)$ from the eigenmode data, defined as $F(\Delta) = N^{-1} \sum_{i=1}^N \Theta(\Delta_i - \Delta)$, where $N$ is the total number of modes, $\Delta_i$ is the detuning of mode $i$ from the band edge, and $\Theta(x)$ is the Heaviside step function. This cumulative distribution represents the fraction of modes with detuning greater than or equal to $\Delta$. We then fit the general stretched-exponential form $F(\Delta) = \exp[-(\Delta/\alpha)^\beta]$ where both $\alpha$ and $\beta$ are treated as free fitting parameters. This approach is crucial because it allows us to test whether the observed spectral tails follow Urbach statistics ($\beta \approx +1$), Lifshitz statistics ($\beta \approx -1/2$), or some other functional form, without imposing any constraint a priori. Taking logarithms gives $-\ln F(\Delta) = (\Delta/\alpha)^\beta$ and hence
\begin{equation}
\ln[-\ln F(\Delta)] = \beta \ln\Delta - \beta\ln\alpha.
\end{equation}
A linear regression of $\ln[-\ln F]$ versus $\ln\Delta$ therefore yields the slope $\beta$ and intercept $b_0 = -\beta\ln\alpha$. From the intercept we obtain
\begin{equation}
\alpha = \exp(-b_0/\beta).
\end{equation}
In this canonical parametrization, $\alpha$ has units of frequency (THz) and is directly the Urbach energy: $\alpha = E_U$. For the ideal Urbach case with $\beta = 1$, the cumulative distribution reduces to $F(\Delta) = \exp(-\Delta/E_U)$.

The fitting procedure optimizes the fitting window to ensure statistical reliability while maximizing linearity. The exponential tail behavior is strictly valid only asymptotically close to the band edge, where disorder-induced band-edge broadening dominates. Therefore, we restrict the fitting window to modes within a maximum detuning $\Delta_{\mathrm{max}}$ from the band edge, chosen to maximize the linearity of the fit as quantified by the coefficient of determination $R^2$. We also exclude the immediate vicinity of the band edge, specifically modes with $F(\Delta) > 0.95$ where finite-size effects and incomplete sampling dominate, as well as modes deep in the tail with $F(\Delta) < 0.05$ where detection limits or finite simulation domains introduce systematic biases. The optimal fitting window thus spans $0.05 < F(\Delta) < 0.95$ and $\Delta < \Delta_{\mathrm{max}}$, where $\Delta_{\mathrm{max}}$ is determined separately for each disorder strength and direction by maximizing $R^2$.

For each disorder configuration ensemble, we compute the cumulative distribution $F(\Delta)$ from all eigenfrequencies across disorder realizations. We then scan candidate values of $\Delta_{\mathrm{max}}$ and perform linear regression on $\ln[-\ln F]$ versus $\ln\Delta$ for each candidate. The value of $\Delta_{\mathrm{max}}$ that maximizes $R^2$ while retaining at least 20 data points for statistical reliability is selected as optimal. From this optimal linear fit, we extract the fitted exponent $\beta$ from the slope and the Urbach energy $\alpha = E_U$ from the intercept via $\alpha = \exp(-b_0/\beta)$, where $b_0$ is the fitted intercept.

\subsection{Numerical results}

Table~\ref{tab:s2_params} presents the complete fitting parameters for all numerical disorder configurations. The fitted exponents cluster consistently around $\beta \approx 1$ across all disorder amplitudes and both directions, confirming Urbach tail behavior. The excellent $R^2$ values (all exceeding 0.96) show that the stretched-exponential form accurately describes the numerical spectral tails.

\begin{table*}[htbp]
\centering
\caption{Complete fitting parameters for numerical simulations (Figure~\ref{fig:numerical_fits}). For each disorder configuration, we report the disorder amplitude $\sigma$, fitted exponent $\beta$ with standard error $\sigma_\beta$, Urbach energy $\alpha = E_U$ (in THz) with propagated error $\sigma_\alpha$, and coefficient of determination $R^2$. The canonical form is $F(\Delta) = \exp[-(\Delta/\alpha)^\beta]$ where $\alpha$ has units of THz.}
\label{tab:s2_params}
\begin{tabular}{cccccc|cccccc}
\hline
\multicolumn{6}{c|}{\textbf{$\parallel$ disorder}} & \multicolumn{6}{c}{\textbf{$\perp$ disorder}} \\
\hline
$\sigma_\parallel$ & $\beta$ & $\sigma_\beta$ & $\alpha$ (THz) & $\sigma_\alpha$ (THz) & $R^2$ & $\sigma_\perp$ & $\beta$ & $\sigma_\beta$ & $\alpha$ (THz) & $\sigma_\alpha$ (THz) & $R^2$ \\
\hline
$0.01a$ & 0.791 & 0.030 & 0.075 & 0.008 & 0.9712 & $0.01a$ & 1.069 & 0.032 & 0.183 & 0.010 & 0.9794 \\
$0.02a$ & 0.988 & 0.050 & 0.260 & 0.020 & 0.9706 & $0.02a$ & 1.510 & 0.038 & 0.401 & 0.010 & 0.9754 \\
$0.03a$ & 1.180 & 0.052 & 0.279 & 0.017 & 0.9608 & $0.03a$ & 1.149 & 0.018 & 0.469 & 0.008 & 0.9848 \\
$0.04a$ & 0.973 & 0.015 & 0.277 & 0.007 & 0.9863 & $0.04a$ & 0.915 & 0.024 & 0.879 & 0.027 & 0.9660 \\
$0.05a$ & 1.267 & 0.014 & 0.390 & 0.005 & 0.9885 & $0.05a$ & 1.039 & 0.023 & 0.698 & 0.017 & 0.9605 \\
\hline
\end{tabular}
\end{table*}

Figure~\ref{fig:numerical_fits} displays the complete fitting analysis for all numerical disorder configurations. The excellent linearity in the log-log plots confirms that the stretched-exponential form with free exponent $\beta$ provides an accurate empirical description of the spectral tails. The fitted exponent values cluster consistently around $\beta \approx 1$, confirming Urbach tail behavior rather than Lifshitz tail behavior across all disorder amplitudes and both directions.

\section{EXPERIMENTAL MEASUREMENTS}

\subsection{Sample growth, fabrication, and disorder implementation}

\begin{figure}[t]
  \centering
  \includegraphics[width=\linewidth]{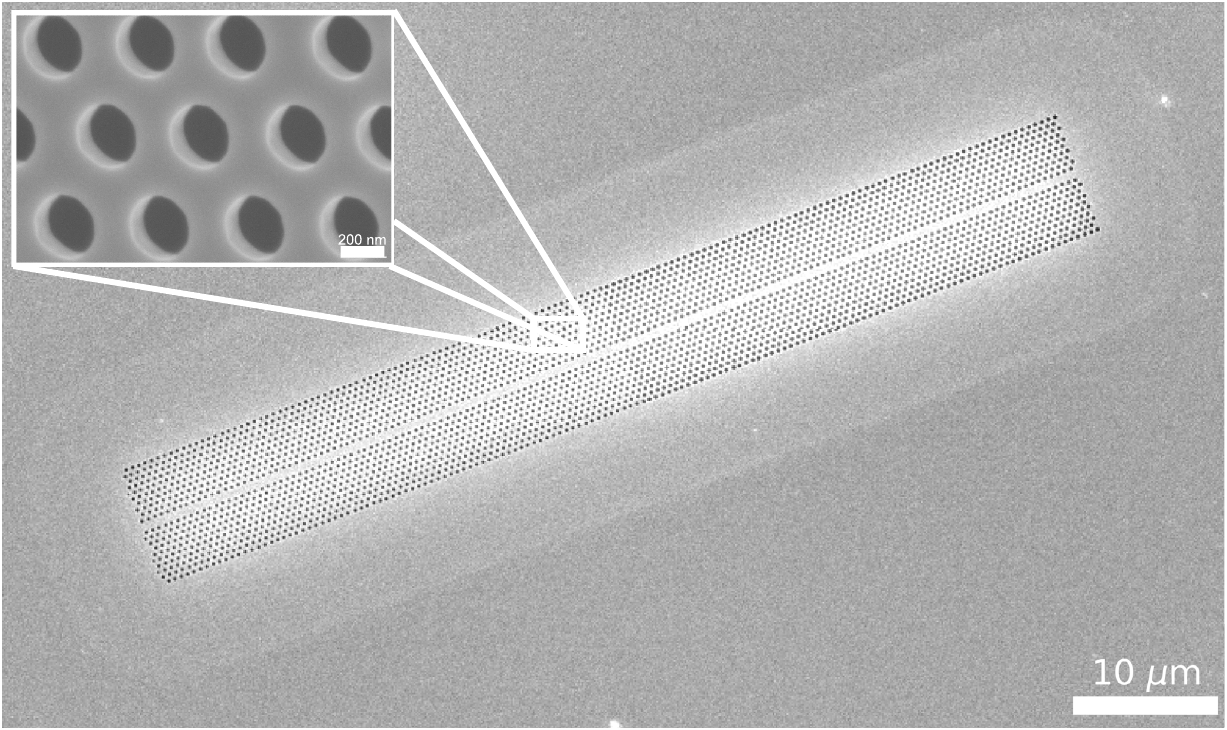}
  \caption{Scanning electron microscope image of a photonic-crystal waveguide. SEM image of a GaAs photonic-crystal waveguide patterned with a triangular lattice of air holes. The waveguide is formed by a single missing row of holes. Inset: close-up of the hole lattice, showing the uniformity of hole size and spacing. Scale bars: 10~$\mu$m (main image) and 200~nm (inset).}
  \label{fig:samples_sem}
\end{figure}

The samples used in this work were grown by molecular beam epitaxy (MBE) on semi-insulating GaAs (100) wafers. After standard thermal desorption at $620\,^{\circ}\mathrm{C}$ under an arsenic overpressure (beam equivalent pressure, BEP, of $1 \times 10^{-5}$~Torr), a 100~nm GaAs buffer layer was grown at $500\,^{\circ}\mathrm{C}$. Subsequently, a 1370~nm-thick Al$_{0.74}$Ga$_{0.26}$As layer was grown at $580\,^{\circ}\mathrm{C}$ using Ga (BEP$_\mathrm{Ga} = 2.69 \times 10^{-7}$~Torr), Al (BEP$_\mathrm{Al} = 1.35 \times 10^{-7}$~Torr), and As (BEP$_\mathrm{As} = 5.98 \times 10^{-6}$~Torr). Finally, a 150~nm GaAs layer was grown on top of the Al$_{0.74}$Ga$_{0.26}$As layer. An in-situ reflection high-energy electron diffraction (RHEED) system was used to calculate the growth rates and monitor the surface quality during epitaxy. Devices were fabricated from the resulting 150~nm-thick GaAs membrane supported by a 1370~nm Al$_{0.74}$Ga$_{0.26}$As sacrificial layer. First, Allresist AR-P~6200.09 electron-beam resist was spin-coated at 3000~rpm for 1~min and baked for 5~min at $170\,^{\circ}\mathrm{C}$. The patterns were exposed at a beam current of 2~nA and a dose of 200~$\mu$C\,cm$^{-2}$, developed in Allresist AR~600-546 for 2~min, rinsed with isopropyl alcohol, and blow-dried with nitrogen. The sample was subsequently descummed in an oxygen plasma for 1~min to remove residual resist from development.

Pattern transfer into the GaAs membrane was performed using chlorine-based inductively coupled plasma reactive-ion etching (ICP-RIE) in a PlasmaTherm Apex SLR system, using BCl$_3$/Ar chemistry with 15~sccm BCl$_3$, 10~sccm Ar, a chamber pressure of 6~mTorr, 500~W ICP power, 25~W RF bias, and an etch time of 2~min 40~s. For etch reproducibility, the ICP chamber was conditioned by running the same etch recipe on a silicon carrier wafer for 15~min prior to processing. After etching, the resist was stripped by soaking the samples in N-methyl-2-pyrrolidone (NMP) at $80\,^{\circ}\mathrm{C}$ for 2~h, followed by 15~min of sonication. The Al$_{0.74}$Ga$_{0.26}$As sacrificial layer was then etched using a 5\% HF solution. Byproducts from the etching processes, including hardened carbon residues from the chlorine-based plasma and insoluble AlF$_3$ formed during HF etching, were removed by sequential immersion in hydrogen peroxide and 22.5\% KOH for 1~min each, with a 1~min deionized water rinse after each solution. The samples were finally dried using CO$_2$ critical-point drying. Figure~\ref{fig:samples_sem} shows a representative scanning electron microscope (SEM) image of a fabricated photonic-crystal waveguide together with a close-up view of the hole lattice, illustrating the overall geometry of the waveguide and the high uniformity of the hole pattern over the full device length. Controlled anisotropic disorder was introduced at the design stage by generating mask files in which each hole position was displaced from its ideal lattice site by a random offset drawn from a Gaussian distribution with zero mean and standard deviation $\sigma$. For $\parallel$ disorder, displacements were applied exclusively along the $\parallel$ coordinate with standard deviation $\sigma_\parallel$, while for $\perp$ disorder, displacements were applied along the $\perp$ coordinate with standard deviation $\sigma_\perp$. The intentional disorder amplitudes span $\sigma \in [0.01a,\,0.05a]$.

\begin{figure*}[b!]
  \centering
  \includegraphics[width=\columnwidth]{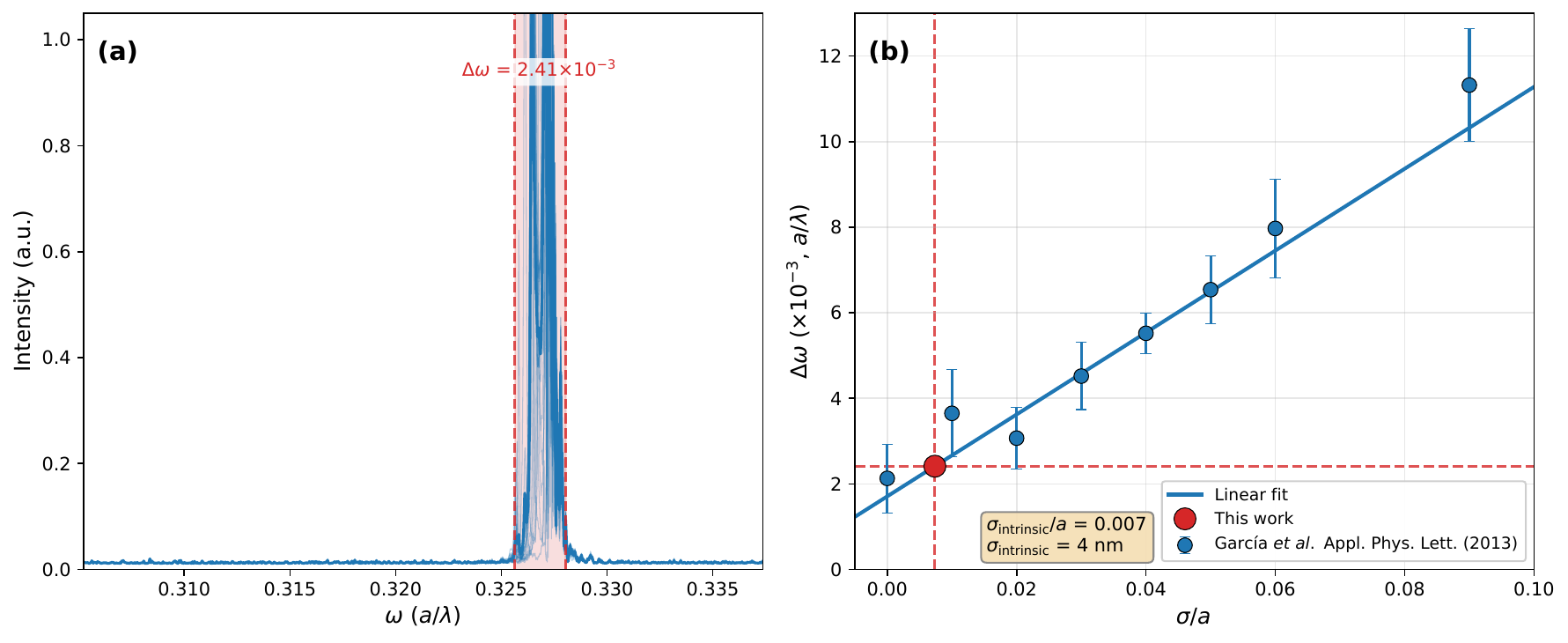}
  \caption{Extraction of intrinsic fabrication disorder. (a) Transmission spectra of a photonic crystal waveguide with no intentionally introduced disorder ($\sigma = 0$), showing 18 measurements at different positions along the waveguide. The shaded region indicates the spectral extent of the Anderson-localized modes, with a tail width $\Delta\omega = 2.41 \times 10^{-3}$ ($a/\lambda$). (b) Calibration of tail width versus positional disorder from Garc\'ia \emph{et al.}~\cite{garcia2013}. The red star indicates the measured $\Delta\omega$ from (a), yielding an intrinsic fabrication disorder of $\sigma_{\mathrm{intrinsic}}/a = 0.007$ ($\sigma_{\mathrm{intrinsic}} = 4$~nm for $a = 500$~nm).}
  \label{fig:intrinsic_disorder}
\end{figure*}

\subsection{Intrinsic fabrication disorder}

\begin{figure*}[tb]
  \centering
  \includegraphics[width=\textwidth]{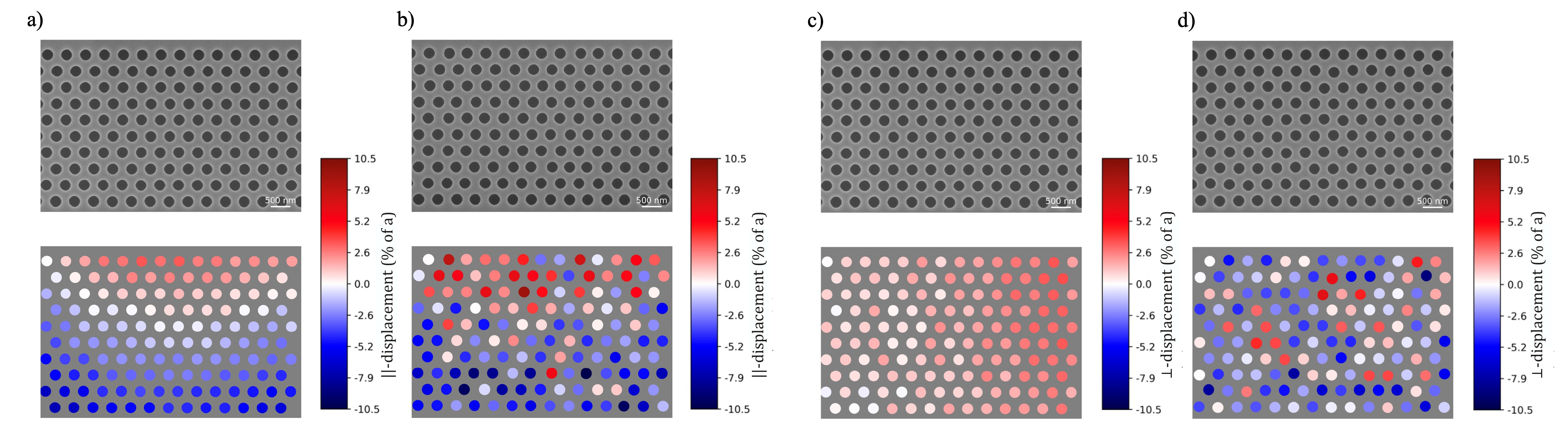}
  \caption{Real-space extraction of positional disorder from SEM images. Top panels show representative SEM images; bottom panels show the corresponding normalized positional displacements obtained by fitting air-hole centers and referencing an ideal hexagonal lattice. (a,b) $\parallel$ disorder: (a) intrinsic $\parallel$ displacement $\Delta_\parallel/a$, (b) engineered $\parallel$ disorder with $\sigma_\parallel = 0.03a$. (c,d) $\perp$ disorder: (c) intrinsic $\perp$ displacement $\Delta_\perp/a$, (d) engineered $\perp$ disorder with $\sigma_\perp = 0.03a$. The comparison demonstrates that intentional disorder dominates over intrinsic fabrication disorder and remains strongly anisotropic.}
  \label{fig:disorder_maps}
\end{figure*}

To quantify the intrinsic fabrication disorder present in our photonic-crystal waveguides, we analyze the spectral width of the Anderson-localized mode tail in samples with no intentionally introduced disorder ($\sigma =0$). Figure~\ref{fig:intrinsic_disorder}(a) shows transmission spectra measured at 18 positions along the waveguide, revealing a distribution of localized modes near the band edge. From the minimal and maximal resonance frequencies we extract a tail width of $\Delta\omega = 2.41\times 10^{-3}$ in normalized units ($a/\lambda$). To translate this tail width into an equivalent positional disorder, we use the calibration established by Garc\'ia \emph{et al.}~\cite{garcia2013}, who measured $\Delta\omega$ as a function of deliberately introduced hole-position disorder in photonic crystal waveguides fabricated on the same material platform. The calibration data, reproduced in Fig.~\ref{fig:intrinsic_disorder}(b) together with a linear fit, provides a direct mapping between normalized tail width and relative disorder amplitude $\sigma/a$. Projecting our measured $\Delta\omega$ onto this calibration yields an intrinsic positional disorder of $\sigma_{\mathrm{intrinsic}}/a = 0.007$, corresponding to $\sigma_{\mathrm{intrinsic}} \approx 4$~nm for the lattice constant used here ($a=500$~nm). This intrinsic value is well below the engineered disorder amplitudes explored in this work ($\sigma=0.01a$–$0.05a$, or 5–25~nm), confirming that the controlled disorder significantly exceeds the fabrication background.

In addition to spectral signatures, intrinsic fabrication disorder can be directly visualized in real space by analyzing scanning electron microscope (SEM) images of the photonic-crystal lattice. The contours of individual air holes are detected and fitted with minimum enclosing circles, whose centers define the actual hole positions. These positions are referenced to an ideal hexagonal lattice with the same lattice constant, using a single reference hole to fix the global origin. The resulting positional deviations are decomposed into components parallel ($\Delta_\parallel$) and perpendicular ($\Delta_\perp$) to the waveguide axis and normalized to the lattice constant $a$. Figure~\ref{fig:disorder_maps} shows representative real-space maps of the normalized positional disorder extracted using this procedure along the $\parallel$ and $\perp$ directions. The comparison demonstrates that intentional disorder dominates over intrinsic fabrication disorder and remains strongly anisotropic.

\subsection{Optical reflection measurements}

Optical reflection measurements are performed using a tunable continuous-wave laser (Santec TSL-570, 1480-1640 nm tuning range) covering the wavelength range 1520-1630 nm (corresponding to 184-197 THz). Light is evanescently coupled into the photonic crystal waveguide using a tapered optical fiber loop placed in contact with the structure along the waveguide axis~\cite{arregui2021}. The reflected light is collected by the same tapered fiber loop and detected using an InGaAs photodetector (Thorlabs PDA20CS2, 800-1700 nm spectral range, 11 MHz bandwidth). The laser wavelength is scanned with a step size of 10 pm, and at each wavelength we measure the reflected optical power. Evanescent coupling to resonant cavity modes appears as sharp spectral peaks in the reflected optical signal. By repositioning the fiber loop along different sections of the waveguide and repeating reflection measurements, we probe Anderson-localized modes at different spatial locations. Each localized resonance appears as a sharp peak in the reflection spectrum. We fit each resonance with a Lorentzian lineshape to extract the resonant frequency $\omega_i$ and quality factor $Q_i$. The detuning of each mode from the band edge is computed as $\Delta_i = \omega_c - \omega_i$ where $\omega_c$ is determined from the onset of the band edge in the reflection spectra.

\subsection{Experimental fitting procedure}

The experimental data consists of resonant frequencies extracted from reflection measurements at multiple spatial positions along the waveguide. For each sample (characterized by disorder direction and amplitude), we compile all detected resonances into a frequency histogram. From this histogram we construct the cumulative distribution $F(\Delta)$ exactly as in the numerical analysis: $F(\Delta) = N^{-1} \sum_{i=1}^N \Theta(\Delta_i - \Delta)$ where $N$ is the total number of detected resonances and $\Delta_i = \omega_c - \omega_i$ is the detuning of resonance $i$ from the band edge.

We then apply the identical fitting procedure used for the numerical data. We fit the stretched-exponential form $F(\Delta) = \exp[-(\Delta/\alpha)^\beta]$ with both $\alpha$ and $\beta$ as free parameters. The fitting window is optimized by scanning candidate values of $\Delta_{\mathrm{max}}$ and selecting the value that maximizes the coefficient of determination $R^2$ while retaining at least 20 detected resonances for statistical reliability. The fitting window is constrained to $0.05 < F(\Delta) < 0.95$ to exclude regions where detection is incomplete or sampling is insufficient. From the optimal linear fit of $\ln[-\ln F]$ versus $\ln\Delta$, we extract the fitted exponent $\beta$ from the slope and the Urbach energy $\alpha = E_U$ from the intercept via $\alpha = \exp(-b_0/\beta)$.

\subsection{Experimental results}

\begin{figure*}[t!]
  \centering
  \includegraphics[width=\textwidth]{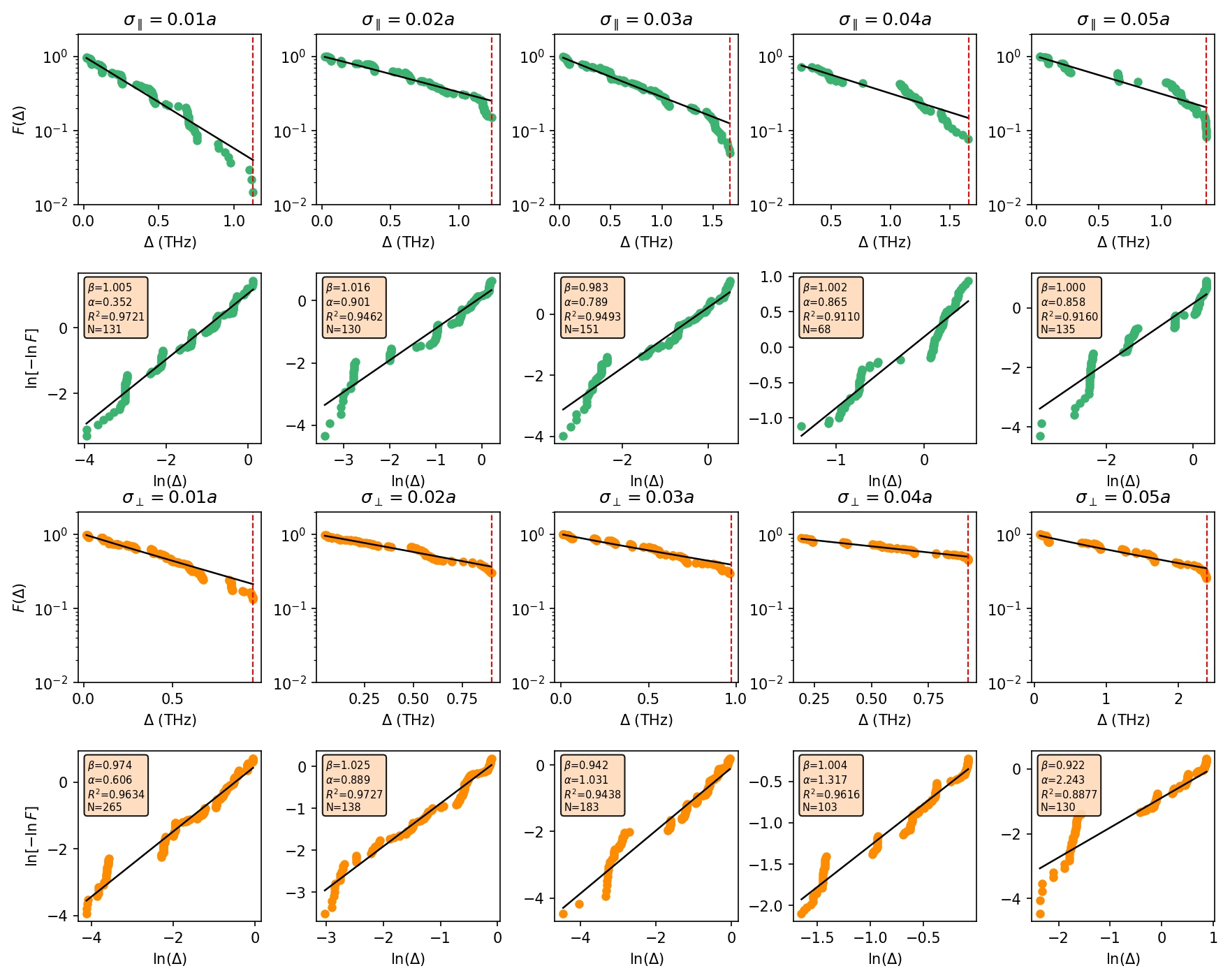}
  \caption{Experimental fitting analysis for directional disorder. The layout mirrors Figure~\ref{fig:numerical_fits}: 4 rows and 5 columns, where columns correspond to disorder amplitudes ($\sigma_\parallel$ or $\sigma_\perp = 0.01a$ through $0.05a$). Rows 1-2 show $\parallel$ disorder: row 1 displays $F(\Delta)$ versus $\Delta$ (THz), and row 2 displays the linearized fit $\ln[-\ln F]$ versus $\ln\Delta$. Rows 3-4 show $\perp$ disorder with the same layout. Colored points (teal for $\parallel$ disorder, orange for $\perp$ disorder) show the optimized fitting region. Black lines show the linear fits. The disorder amplitudes represent the designed values programmed into the e-beam mask; the actual total disorder includes intrinsic fabrication disorder of $\sim0.01a$-$0.02a$. The good linearity of the experimental data confirms that the stretched-exponential form accurately describes the experimental spectral tails. The fitted exponent values cluster around $\beta \approx 1$ for all samples, in quantitative agreement with the numerical simulations and confirming that near-band-edge disorder-induced states in photonic crystal waveguides exhibit Urbach tail behavior, not Lifshitz tail behavior.}
  \label{fig:experimental_fits}
\end{figure*}

Table~\ref{tab:s3_params} summarizes the experimental fitting results for samples with intentional disorder along both $\parallel$ and $\perp$ directions. The disorder amplitudes listed represent the total effective disorder including both intentional and intrinsic contributions. The fitted exponents show $\beta \approx 1$ for all measured configurations, consistent with Urbach tail behavior. The coefficient of determination exceeds $R^2 > 0.89$ for all cases, confirming good linearity within the optimized fitting windows despite the additional complexity of experimental measurements compared to idealized simulations.

\begin{table*}[htbp]
\centering
\caption{Complete fitting parameters for experimental measurements (Figure~\ref{fig:experimental_fits}). For each sample, we report the disorder amplitude $\sigma$, fitted exponent $\beta$ with standard error $\sigma_\beta$, Urbach energy $\alpha = E_U$ (in THz) with propagated error $\sigma_\alpha$, and coefficient of determination $R^2$. The canonical form is $F(\Delta) = \exp[-(\Delta/\alpha)^\beta]$ where $\alpha$ has units of THz.}
\label{tab:s3_params}
\begin{tabular}{cccccc|cccccc}
\hline
\multicolumn{6}{c|}{\textbf{$\parallel$ disorder}} & \multicolumn{6}{c}{\textbf{$\perp$ disorder}} \\
\hline
$\sigma_\parallel$ & $\beta$ & $\sigma_\beta$ & $\alpha$ (THz) & $\sigma_\alpha$ (THz) & $R^2$ & $\sigma_\perp$ & $\beta$ & $\sigma_\beta$ & $\alpha$ (THz) & $\sigma_\alpha$ (THz) & $R^2$ \\
\hline
$0.01a$ & 1.005 & 0.015 & 0.354 & 0.008 & 0.9721 & $0.01a$ & 0.974 & 0.012 & 0.598 & 0.009 & 0.9634 \\
$0.02a$ & 1.016 & 0.021 & 0.902 & 0.019 & 0.9462 & $0.02a$ & 1.025 & 0.015 & 0.892 & 0.011 & 0.9727 \\
$0.03a$ & 0.983 & 0.019 & 0.786 & 0.017 & 0.9493 & $0.03a$ & 0.942 & 0.017 & 1.033 & 0.021 & 0.9438 \\
$0.04a$ & 1.002 & 0.039 & 0.865 & 0.019 & 0.9110 & $0.04a$ & 1.004 & 0.020 & 1.316 & 0.015 & 0.9616 \\
$0.05a$ & 1.000 & 0.026 & 0.858 & 0.024 & 0.9160 & $0.05a$ & 0.922 & 0.029 & 2.402 & 0.108 & 0.8877 \\
\hline
\end{tabular}
\end{table*}

Figure~\ref{fig:experimental_fits} displays the complete experimental fitting analysis. The layout mirrors that of Fig.~\ref{fig:numerical_fits}. The good linearity of the experimental data confirms that the stretched-exponential form accurately describes the experimental spectral tails. The fitted exponent values cluster around $\beta \approx 1$ for all samples, in quantitative agreement with the numerical simulations. The experimental data in Fig.~\ref{fig:experimental_fits} exhibit clustering into small bunches, a feature absent in the numerical simulations (Fig.~\ref{fig:numerical_fits}), which we attribute to the discrete spatial sampling along the waveguide. Despite this clustering, the global fit correctly recovers the Urbach exponent $\beta \approx 1$, confirming the robustness of the underlying Urbach statistics.

\section{CONCLUSIONS}

This supplementary material establishes a rigorous framework for distinguishing Urbach and Lifshitz tail regimes in photonic-crystal waveguides. By treating the exponent $\beta$ as a free fitting parameter in the stretched-exponential form $F(\Delta) = \exp[-(\Delta/\alpha)^\beta]$, we show that near-band-edge disorder-induced states exhibit Urbach tail behavior ($\beta \approx +1$) rather than Lifshitz tail behavior ($\beta \approx -1/2$). The Urbach energy $\alpha = E_U$ provides a direct quantification of the disorder-induced band-edge broadening. Our comprehensive analysis of both numerical simulations and experimental measurements across multiple disorder amplitudes and directions consistently yields $\beta \approx 1$, conclusively confirming the universality of the photonic Urbach tail in disordered photonic-crystal waveguides.

\end{document}